\documentclass[]{aa} % for a referee version
\usepackage{graphicx}
\usepackage{txfonts}
\usepackage{threeparttable}
\usepackage{xtab}
\usepackage{array}
\usepackage{natbib}
\begin{document}
   \title{Constraining the thermal history of the Warm--Hot
   Intergalactic Medium}

\author{L. Zappacosta\inst{1}, R. Maiolino\inst{2}, 
A. Finoguenov\inst{3}, F. Mannucci\inst{4},
R. Gilli\inst{2},\\ A. Ferrara\inst{5}}

\institute{Dipartimento di Astronomia e Scienza dello Spazio, Largo
E. Fermi 2, I-50125 Firenze, Italy, 
\and Osservatorio Astrofisico di Arcetri Largo E. Fermi 5, I-50125 Firenze, Italy 
\and Max-Planck-Institut f\"ur extraterrestrische Physik, Giessenbachstra\ss e, D-85748 Garching, Germany
\and Istituto di Radioastronomia, Sezione di Firenze - CNR Largo E.Fermi 5, I-50125 Firenze Italy 
\and SISSA/International School for Advanced Studies Via Beirut, 4
34014 Trieste, Italy
}
   \offprints{L. Zappacosta, \email{zappacos@nabhas.ps.uci.edu}}

   \authorrunning{Zappacosta et al.}
   \titlerunning{Constraining the thermal history of the WHIM}

   \date{Received ; accepted }

   \abstract{We have identified a large-scale structure traced by galaxies at z=0.8,
   within the Lockman Hole, by means of multi-object spectroscopic observations. By
   using deep XMM images we have investigated the soft X-ray emission from
   the Warm-Hot Intergalactic Medium (WHIM) expected to be associated with  
   this large-scale structure and we set a tight upper limit to its flux in the
   very soft 
   0.2--0.4~keV band. The non-detection requires the WHIM at these redshifts
   to be cooler than 0.1~keV.  Combined with the WHIM emission detections at
   lower redshift, our result indicates that the 
    WHIM temperature is rapidly decreasing with redshift, as expected in
   popular cosmological models.

   }

   \maketitle
%
%_________________________________________________________________

\section{Introduction}

Both observations and cosmological models suggest that most of the baryonic
matter is located in the intergalactic medium (IGM).  Cosmological models
have also shown that the evolution of the baryonic matter is driven by
progressive gravitational heating in the potential field of dark matter
filaments \citep{cen,dave}. In particular, at high redshifts ($z>$2--3) the
baryonic gas is relatively cold ($T< 10^5$~K), and is identified with
Ly$\alpha$ absorbers along the line of sight of quasars, while at lower 
redshifts ($z<$1) the shocks due to the infall of the gas on the dark 
matter filaments (traced by the regions of high galaxy number densities)
gradually heat the gas to temperatures in the range 
T$\sim 10^5-10^7$~K. 

The identification of such Warm-Hot Intergalactic Medium (WHIM) in the local
universe has received growing interest in the last few years. The detection of
absorption lines from highly ionized species ($\rm{O\,VI}$,
$\rm{O\,VII}$, $\rm{O\,VIII}$), both in the
UV and in the soft X-rays, has allowed to unambiguosly identify WHIM along
the line of sight of a few bright quasars \citep{nicastro,mathur}. The
detection of emission due to WHIM is quite challenging. Indeed, the WHIM is
expected to emit weak and diffuse radiation mostly in the softest X-ray
bands ($\la \,1keV$), where both the Galactic absorption and the foreground
emission from the Local Hot Bubble (LHB) and Milky Way Halo are
strong. Nevertheless, several
independent detections of WHIM emission were obtained by detailed analysis
of soft X-ray maps (ROSAT and XMM) in regions characterized by galaxy
overdensities and by the spectral analysis of clusters of galaxies and of
their surroundings \citep{wang,soltan,zappacosta,zappacosta2,kaastra,finoguenov}.

All WHIM detections discussed above have been obtained for gas in the local
universe or at low redshift. The most distant WHIM emission detected so far
was obtained at z$\sim$0.45 by \citet{zappacosta}. At higher redshift the
detection is more difficult because of both technical and physical
reasons.  Indeed, the lack of bright, high redshift quasars
prevents the detection of WHIM features in absorption, while the thermal
cutoff is redshifted to lower energies making more difficult to detect the
WHIM emission even in the soft X-rays. An additional issue is that,
according to cosmological models, the WHIM should be cooler at higher
redshift, implying a lower ionization state of the gas (i.e. lower optical
depth of high ionization absorbers) and a thermal cutoff further moved
to lower energies.  Yet, it would be
most useful to obtain some constraints on the WHIM properties at high
redshift since, when compared with the WHIM detections in the local
universe, it would provide contraints on the cosmological models of the
evolution of baryons.

To pursue the latter goal we started a detailed investigation of the Lockman
Hole, which is one of the fields where the Galactic absorption is
minimum ($N_H \sim 5.6\times10^{19} \,\rm{cm^{-2}}$)
and where deep X-ray observations have been obtained \citep{hasinger4,hasinger3,hasinger}.
We have searched for
superstructures by analyzing the redshift distribution of sources already
identified in this field and found 8 sources in a narrow redshift range at
about z$\sim$0.8, located within a region of about 20 arcmin. We obtained
multi-object spectroscopy in the same area and, as discussed in
Sect. \ref{opt_obs}, we have confirmed the presence of a superstructure at
z$\sim$0.8. Then we have analyzed an XMM map of the LH in the softest band
and, as discussed in Sect. \ref{X_obs}, we have obtained tight constraints
on the possible diffuse emission due to WHIM associated with the large-scale
structure. In Sect. \ref{discuss} we discuss these observational
constraints on the evolution of the WHIM with redshift.

\section{Optical observations: detection of a large-scale
structure at z$\sim$0.8}\label{opt_obs}

As mentioned in the Introduction, an analysis of the redshift distribution
of the sources previously identified in the Lockman Hole (most of which are
   \begin{figure}
   \centering
   \includegraphics[angle=270, width=0.45\textwidth]{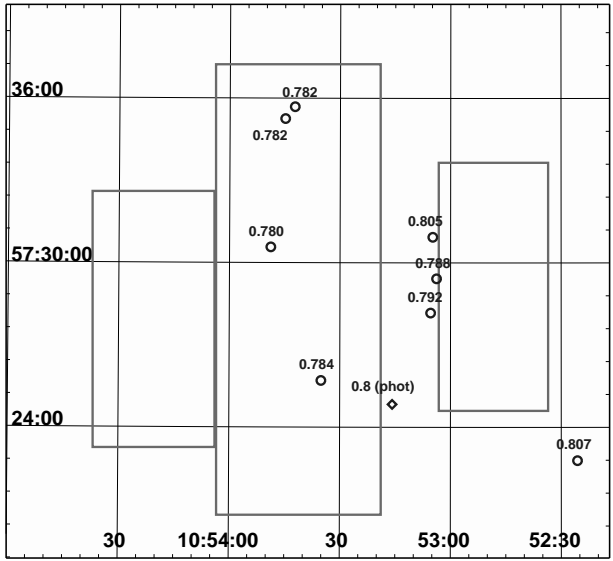}
      \caption{Spatial distribution of objects reported
in the literature in the narrow redshift
range 0.78$<$z$<$0.81 within the Lockman
Hole (see Table~\ref{arch_tab}). Each object is labelled with its redshift.
The object marked with a diamond has a
photometric redshift. Boxes indicate the regions
covered by our spectroscopic survey
with 10 masks.}
      \label{archiv_map}
   \end{figure}
optical counterparts of X-ray sources detected by ROSAT and XMM) has
revealed the existence of 8 sources in the narrow redshift range
0.780--0.807 ($\sim 74 \,\rm{Mpc}$\footnote{In the whole paper we assume a
  cosmology with $\Omega_{m} = 0.3$, $\Omega_{\Lambda} =
  0.7$ and $\rm{H_{0}} = 70 \,km\,s^{-1}\,Mpc^{-1}$ and all the 
distances will be expressed in the comoving rest frame.}). These sources (listed in table \ref{arch_tab})
are marked with a circle in Fig. \ref{archiv_map} and are located within a
region of about 20~arcmin. In the same region we have found another object
(marked with a diamond) with a photometrically estimated redshift of
$0.8\pm0.1$. Note that 4 of such objects (distributed along the N--S
direction) have redshifts in the narrower interval 0.780--0.784. This is
strongly suggestive of the existence of a large-scale structure at this
redshift in this region of the Lockman Hole.

To confirm this tentative indication we have obtained multi-object
spectroscopy of 215 galaxies in the same area. We used the
multi-object spectroscopic mode (MOS) of the optical spectrometer 
DOLORES, at the Telescopio
Nazionale Galileo (TNG), with the LR--R grating, which covers
the 4470--10360 \AA \ range at a resolution of $11\AA$. 
This spectral range allows the identification of
H$\beta$+[OIII] and [OII] at z$\sim$0.8.
\begin{table*}
 \begin{center}
\caption{Objects in the narrow redshift range
0.78$<$z$<$0.81 located within the Lockman Hole
reported in the literature.}
  \begin{tabular}{lllclr}
\hline
\hline
 {\bf Object}          &  {\bf RA(J2000)}  & {\bf DEC(J2000)}   & {\bf Type}       & \multicolumn{1}{c}{{\bf z}}         & {\bf Reference} \\
\hline
RDS 117Q               &  10h 53m 48.8s & $+57^{\circ} 30^{\prime} 34^{\prime\prime}   $ & AGN          & 0.780     & \citet{lehmann}\\
$[$HGG98$]$ 5          &  10h 53m 44.9s & $+57^{\circ} 35^{\prime} 15^{\prime\prime}   $ &   Galaxy          & 0.782     & \citet{hasinger2}\\
$[$HGG98$]$ 8          &  10h 53m 42.3s & $+57^{\circ} 35^{\prime} 41^{\prime\prime}   $ &   Galaxy          & 0.782     & \citet{hasinger2}\\
RX J105335.1+572542    &  10h 53m 35.1s & $+57^{\circ} 25^{\prime} 42^{\prime\prime}   $ & QSO          & 0.784     & \citet{mainieri}\\
RX J105303.9+572925    &  10h 53m 03.9s & $+57^{\circ} 29^{\prime} 25^{\prime\prime}   $ & QSO          & 0.788     & \citet{mainieri}\\
$[$MBH2002$]$ 41       &  10h 53m 05.4s & $+57^{\circ} 28^{\prime} 10^{\prime\prime}   $ & X--ray source& 0.792     & \citet{mainieri}\\
$[$FFH2002$]$ 105      &  10h 53m 15.80s& $+57^{\circ} 24^{\prime} 50.0^{\prime\prime} $ & X--ray source& 0.8 (phot)& \citet{fadda}\\
LOCK-6cm J105304+573055&  10h 53m 04.83s& $+57^{\circ} 30^{\prime} 55.9^{\prime\prime} $ & Radio source & 0.805     & \citet{ciliegi}\\
RX J105225.3+572246    &  10h 52m 25.3s & $+57^{\circ} 22^{\prime} 46^{\prime\prime}   $ & AGN          & 0.807     & \citet{mainieri}\\
\hline
  \label{arch_tab}
  \end{tabular}
 \end{center}
\end{table*}
The observations were performed during four nights in March 2003.

We selected two samples of galaxies within the subarea
of interest of the Lockman Hole. In particular, we selected a {\it shallow sample},
made of galaxies in the magnitude range $\rm 20.0\leq R<21.3$, and
a {\it deep sample}, containing galaxies in the magnitude range $\rm 21.3\leq R<22$.
The {\it shallow sample} was observed with seven masks and with an
integration of 1.5 hours for each mask, while the {\it deep sample} was
observed with three masks and with an integration of 3 hours per mask.
Most of the masks (7 of them) were located along the N--S direction traced by
the 4 sources at z=0.78, while three masks were located to
the west and to the east to check possible extensions of the putative large
scale structure. The location of the various masks is shown by the boxes in
Fig. \ref{archiv_map}.

The MOS spectra were reduced following the standard
   \begin{figure}[t]
   \centering
   \includegraphics[angle=0, width=0.5\textwidth]{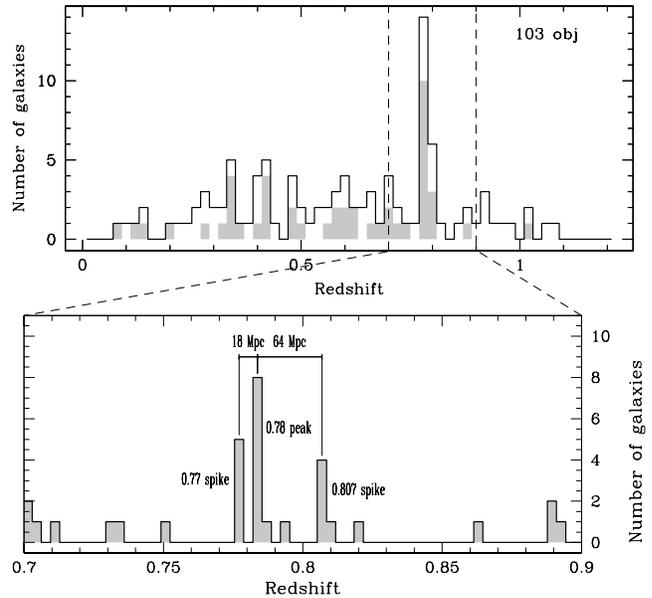}
      \caption{Upper panel: redshift distribution of the objects for which
        we have obtained high quality (shaded histogram) and low quality
        (hollow histogram) spectroscopic redshift measurements. Lower panel:
	zoom of the redshift distribution around the three main
        concentrations of objects.}
      \label{hist_z}
   \end{figure}
threads (dark and bias subtraction, flat fielding, wavelength
calibration). The {\em deep sample} was observed using a dithering of 6
arcsec along the slit axis to enable a better subtraction of sky lines and
an easiest detection of the weak emission lines.  
We observed 215 sources in total. 
We measured redshifts for 103 objects (see Appendix \ref{catalog} for 
the catalog) with typical random errors of
$\pm0.002$, as inferred by the uncertainty on the gaussian fit to the
emission lines\footnote{ The only source of possible systematic errors is
    spectral calibration. However, such systematic errors are smaller
    than the estimated random errors.}.
For 47 sources the 
redshift could be determined unambiguously thanks to the detection of two or
more emission lines. We assigned unambiguously the redshift also in case of
detection of only one strong emission line identified as [OII] line at high
redshift \footnote{Alternatively it would be the H$\alpha$ of local objects
(z$\sim0$), but in the latter case the objects should be extended and bright,
unlike the ones in our samples.} (in Appendix
  \ref{catalog} the quality of these redshifts is marked as ``high'').  
For other 56 objects the redshift
determination was less secure, based on line detections with low signal to
noise ratio  (in Appendix
  \ref{catalog} the quality of these redshifts is marked as ``low'').
For 112 sources the spectrum was too weak and without bright lines, and we
could not recover any information on their redshifts. 

\begin{figure*}
   \centering
   \includegraphics[angle=270, width=0.95\textwidth]{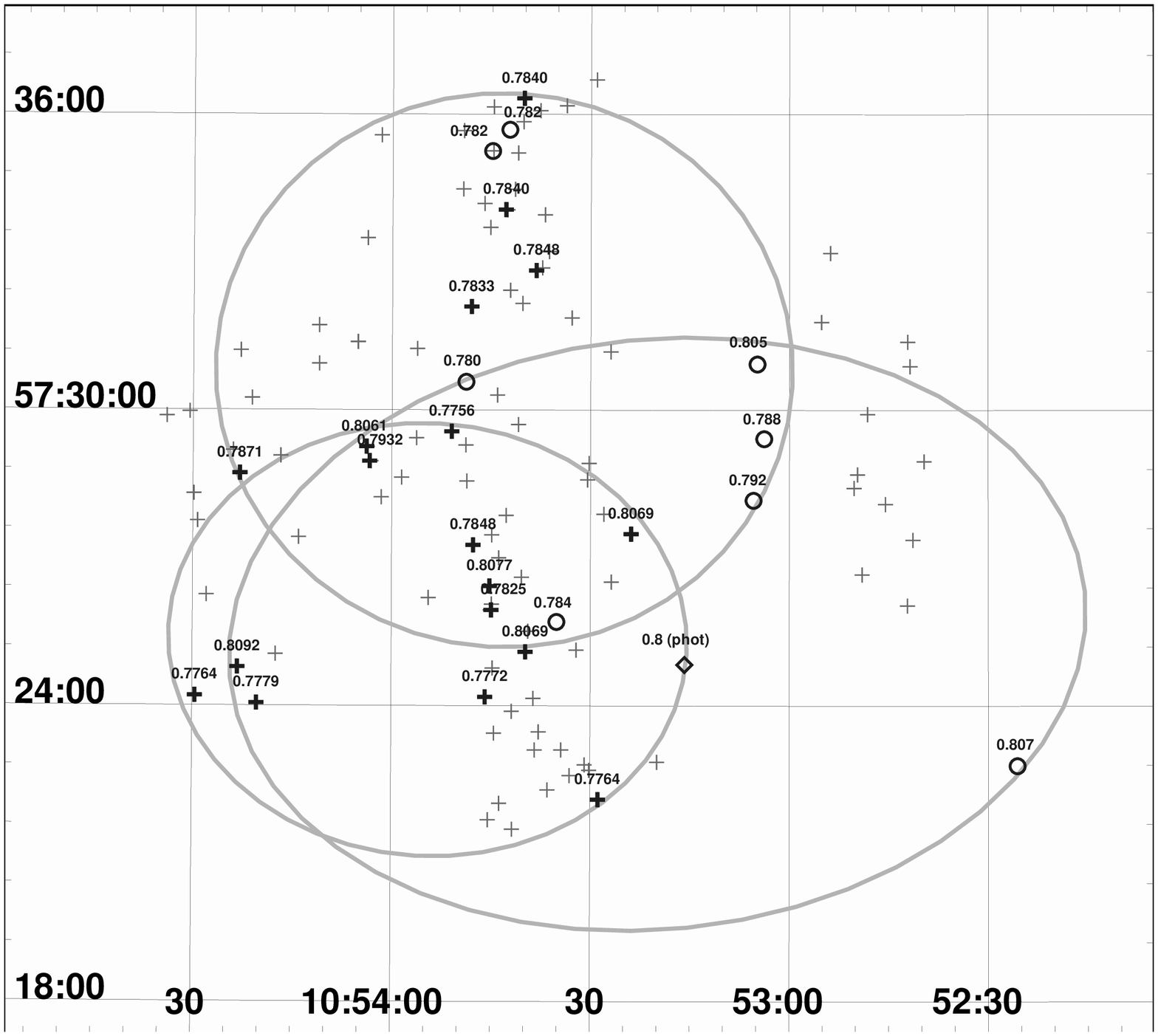}
      \caption{ Spatial distribution of the various spectroscopically
        identified sources and in particular of those at the
        redshift corresponding to the spikes at z$\approx$0.8.
        Crosses mark the position of all the galaxies with redshifts
        measured by us (the thick ones are the sources belonging to
        the superstructure). The circles and the diamond are the archival
        objects at the same redshift of the spike (as in 
	Fig. \ref{archiv_map}). For all the
        galaxies belonging to the superstructure we have reported
        their measured redshift. 
        The large ellipses tentatively indicate the regions
        which may be mostly populated by the three 
        subspikes.
}
      \label{struct}
   \end{figure*}

The redshift
distribution is shown in the top panel of Fig. \ref{hist_z}, where a
prominent peak ($6 \sigma$ significant with respect to the mean histogram
level) is seen at the expected redshift of z=0.78, demonstrating
beyond any doubt the presence of a large-scale structure at this redshift in
this region.  As for the archival objects, this spike shows two distinct
sub-concentrations one at z$\sim 0.78$ and one at z$\sim 0.807$. The
former is further splitted into a main peak at z$\sim0.784$ and 
a nearby spike at
z$\sim0.776$, as shown in the bottom panel. Fig. \ref{struct} shows the
distribution projected on the sky of the sources for which the redshift could be
determined, and in particular for those in the redshift spikes. 
Although this redshift survey is not complete, our data
suggest that these structures at slightly different redshifts tend also to be 
distributed in different regions of the sky. In particular,
galaxies on the red tail of the main spike, and specifically at 
z$\sim0.776$ are located in
the southern part of the field, while galaxies in the main spike at
z$\sim0.784$ are preferentially distributed in the northern part. The
distribution of the galaxies in the farther spike at z$\sim0.807$ overlaps
with the previous two, but the presence of another source at 0.807 (the
one located most to the west in Fig.\ref{struct}),
previously identified by \citet{lehmann},
suggests that this substructure extends towards the west. We have
tentatively encircled the three main substructures with three ellipses in
Fig.\ref{struct}.
On the whole the superstructure outlined by our survey and the
archival objects cover a region
of $\sim 7.5 \,\rm{Mpc}$ (at the mean redshift z=0.791). This could be considered as a lower limit to
the dimension of the structure because its size is limited by the
extent of our observations.
However, we note that other spectroscopic surveys covering the entire
XMM and ROSAT-HRI fields have not found additional objects in the
narrow redshift range outside the area considered by us, suggesting
that the large-scale structure is not further extended at least in the
XMM and ROSAT-HRI fields.

\section{X-ray data: constraints on the WHIM emission}\label{X_obs}

Once demonstrated that large-scale structures at high redshift exist within
the Lockman Hole, we have then investigated whether the associated WHIM emission
could be detected. Such investigation requires sensitive maps at soft
energies E$<$0.5~keV, both because of the low temperatures typical of 
WHIM and because of a significant Doppler shift of the emission
to lower energies. 
Several deep X-ray
observations have been carried out in the Lockman Hole area, most importantly
with ROSAT and XMM-Newton \citep{hasinger4,hasinger3}.

ROSAT maps would have the required field of view and sensitivity at low
energies to properly constrain the presence of diffuse warm gas at high
redshifts. However, the region which we are investigating is also
characterized by a high density of X-ray point sources (a fraction of which
belonging to the large-scale structure at z=0.78) and by a few clusters of
galaxies \citep{hasinger2}. %{\bf (Szokoly et al. 2003??)}. 
After the subtraction of the instrumental background and the point
source removal through the procedures described in \citet{zappacosta},
we find diffuse emission coincident with the superstructure.
However, the relatively extended 
wings of the ROSAT PSF from the point sources probably contribute
significantly to the diffuse emission.
Indeed, this apparently extended emission elongated in the direction N--S is
also found in the
harder maps (R45 and R67). The spectral shape of such extended emission
is the same as for the point sources, associating this emission with
residual wings of point sources as well as a
possible detection of unresolved AGNs associated with the large-scale structure
\citep[e.g.][]{gilli}. The observed hardness of the emission is also at
variance with what found in other fields where diffuse emission has a much
softer spectrum \citep[e.g.][]{zappacosta}. The presence of such
N-S unresolved AGN emission further suggests the presence of an
overdense region of galaxies in this area of the Lockman Hole.

Chandra has a much better angular resolution which allows the removal of the
contribution from point sources with a much higher accuracy. 
However, its sensitivity drops
drastically at energies E$<$0.5~keV,  preventing us to
study the level of soft diffuse X-ray emission.
The small field of view of ACIS-S (the Chandra chip
which has higher sensitivity in the soft band
than ACIS-I) is also problematic to detect extended
emission.

XMM has the appropriate compromise between angular resolution, good
sensitivity in the softest X-ray band at E$<$0.5~keV, and extension of
the field of view. The Lockman Hole has
been subject of deep observations with XMM
\citep{hasinger3,hasinger}. 
In particular, the 100~ksec observation obtained by \citet{hasinger3} 
was performed
 with the ``thin'' filter, which
allows the detection of photons down to 0.2~keV.
Additional 800~ksec of integration \citep{hasinger} were obtained
 with the ``medium'' filter, which absorbs
photons with E$<$0.5~keV. The latter observation cannot
be used to constrain the WHIM emission because of
its energy cutoff at 0.5~keV, however it can be
efficiently used to subtract spurious contribution
to the diffuse emission by hard sources, as we shall
discuss later on. In the following we will focus
on the analisys of the XMM data taken with the
``thin'' filter.

The major difficulty in using the soft 
energy band is the presence of the electronic noise that dominates the
emission \citep{lumb,read}. The spectrum 
of the electronic noise is very stable, and its statistical noise consists
in a number of small-amplitude events occurring during every frame
exposure. There are no fluctuations in a
number of events, so the removal of these has no major influence on the
detection statistics for the X-ray emission. Electronic noise has a similar
spatial distribution for the same detector read-out mode, but its spectrum
varies as a function of frame time and is subject to an energy offset
on the 10
eV scale for each individual pixel. This energy offset applies to all
events, resulting in a decrease of the energy resolution for an extended
source. Investigations of the electronic noise by the EPIC calibration team
at the Max-Planck Institut f\"ur Extraterrestrische Physik showed that it is
possible, by using the shape of the electronic noise, to actually determine the
energy offset in each pixel for each observation and then 
efficiently and accurately 
remove the electronic noise from the event lists. The software and
processing recipes are made available to a general user as a new task,
{\it epreject}, within SAS 6.0 release. In this Paper we
describe the results obtained for a single Lockman Hole pointing as a part
of software testing stage by Konrad Dennerl (see \citet{dennerl} and
{\it epreject} task description). For further details of the
noise removal and a discussion of associated uncertainties see {\it
epnreject} task description. We selected the longest EPIC-pn observation, made
with the ``thin'' filter and detector full frame mode \citep[for a description
of EPIC-pn see][]{struder}, which yields 35~ksec of useful exposure. For
purposes of our analysis, a subtraction of out-of-time events (OOTE) is not
necessary, given the position of bright sources with respect to the 
superstructure and an orientation of CCDs in the selected observation. 
The sensitivity reached by this observation is good enough to set
relatively tight constraints on the presence of diffuse soft X-ray
emission in the region of the Lockman Hole.

\begin{figure}
   \centering
   \includegraphics[angle=270, width=0.45\textwidth]{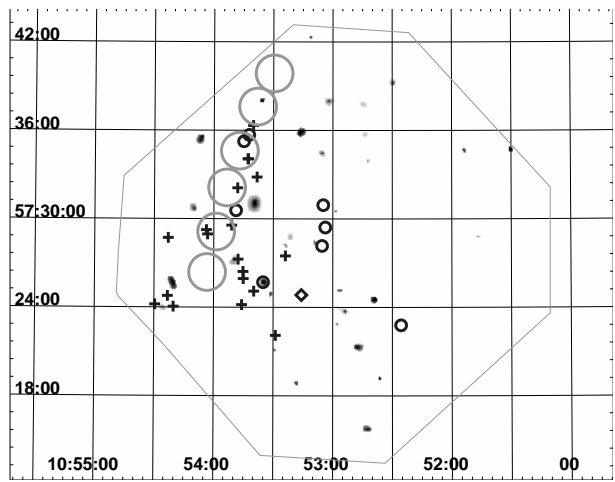}
      \caption{XMM-Newton EPIC-pn wavelet reconstructed image in the
        0.2--0.4 keV band. Crosses, small circles and the diamond are the galaxies belonging
      to the superstructure detected in optical as in Fig. \ref{struct}. Big circles are the regions where we measured the
      residual soft X-ray flux. The thin line shows the position
      of the EPIC-pn camera.}
      \label{xmm_wv}
   \end{figure}

%The 100~ksec observation was reduced
%very carefully with specific attention to any instrumental issue
%which might affect the detection of diffuse emission. In particular,
%the instrumental background, obtained by a long exposure with the shutter
%closed %(courtesy of ???), 
%was removed from each sub-exposure
%before combining them to obtain the final, mosaiced image.
%{\bf (Alexis, you probably want to add something or to provide
%a better description of the reduction steps...).}
The image in the softest available band 0.2--0.4~keV was extracted
and processed through a variable wavelet filter detection algorithm \citep{vikhlinin}
to identify both point and extended sources  
\citep[see][ for details]{zappacosta}. 
We set the wavelet peak detection
threshold to $4\sigma$, and followed the extension of the detected flux
down to $1.7\sigma$. We performed ten
iterations at wavelet kernel scales ranging from $4^{\prime\prime}$  to $4^{\prime}$. 
We started from the smallest scale and removed the detected sources
prior to proceeding with the next larger scale.
``Point sources'' were identified as sources detected with
kernels of size $4^{\prime\prime}$ (central part of the field of view) and $8^{\prime\prime}$ (outer
region, where the PSF is larger).
The resulting wavelet map (the sum of all wavelet orders) is
shown in Fig.~\ref{xmm_wv}.
Several
point sources are detected as well as a few known clusters. However, we
do not find any evidence for diffuse extended emission down to
a limit of 2$\sigma$ of confidence from the high order wavelet
maps, particularly in the region where the large-scale structure at
z$=$0.78 has been detected.

To place limits on the presence of a soft diffuse component associated with
the optical structure, identified above, we have selected a number of
circles along the superstructure traced by galaxies
(see in Fig. \ref{xmm_wv} the big circles).
We masked out from final
analysis all the parts where X-ray emission has been detected in the 0.5--2
keV and in the 2--7.5 keV bands in the full $\sim800$ ksec XMM-Newton 
exposure on the
Lockman Hole, obtained with the ``medium'' filter; 
in this way we could remove instrumental
scattering effect, two clusters located near
the superstructure, as well as the
contribution by unresolved AGNs (which have
a much harder spectrum than the WHIM). 
We then used the 0.2--0.4~keV band to extract the counts from the 
observation with the thin filter. We also selected two 
additional regions, closer to the center of the 
pointing and at the edge of the CCD to estimate both a level of the
foreground and background emission close to the position of our
measurement and to estimate the possible contribution of the induced
background, as due to the soft protons. To do that, we make a reasonable
assumption, based on our best knowledge, 
that the sky components and X-ray foreground
components (such as the Local Hot Bubble, LHB)
have flat distribution on the sky, at least
within the XMM field of view, and thus are
vignetted by the
telescope, while the induced background components have a flat distribution of
their intensity over the detector (e.g. Lumb et al. 2002).

After subtraction of these in-field estimated background components, the
residual soft X-ray flux is $\rm F_{0.2-0.4keV}=1.5\pm 1.4 \times 
10^{-16}~erg~s^{-1}~cm^{-2}~arcmin^{-2}$, that increases by $\sim 30\%$
after the correction for HI absorption.
This is a very marginal detection
and could be explained by statistical fluctuations around the zero
value. Therefore, we consider such a flux as an upper limit to the soft
diffuse X-ray emission in this region.
 It should be noted that the Lockman Hole is the
clearest window for this kind of X--ray studies, being the region with the lowest
Galactic hydrogen column density. This means that a superstructure, as
the one we have found in optical, should clearly show X--ray
diffuse emission due to the collapsed gas in the dark matter potential
well traced by the galaxies, unless the gas temperature is very low. 
This issue will be discussed more in detail in the next section.

\section{Temperature of the WHIM at z$\sim$0.8}\label{discuss}

In this section we discuss the upper limit obtained for the diffuse
X-ray emission and estimate whether it is compatible with the
predictions of the cosmological models and/or constraints on the latter can
be inferred.

In the local universe ($\rm{z}\,<1$) various detections of emission by WHIM
indicate temperatures ranging from $\sim$1~keV down to about
$\sim$0.2~keV
\citep{wang,zappacosta,zappacosta2,soltan,kaastra,finoguenov}.  The lower
temperature limit could be ascribed to observational issues, since the
bulk of the emission from gas with the temperatures below 0.1 keV 
is absorbed even in regions of low Galactic $\rm{N_H}$. Moreover, low 
Oxygen abundance characteristic of
the WHIM \citep{cen,finoguenov} 
%, as well as recently discovered transient heliospheric components (Kunz \&
%Snowden 2004),
 precludes WHIM detection through the OVII emission 
lines. The situation
will change, however, with advent of microcalorimeters with 'large grasp'. 
The alternative technique
of using the OVII and OVIII X-ray absorption 
lines allowed the detection of Local WHIM at much lower temperatures
\citep{nicastro}. For a homogeneous comparison, here
we focus on the properties of the WHIM detected in emission.

We take as a reference
for the local universe the WHIM emission from the Sculptor
supercluster (z=0.1)
reported in \citet{zappacosta2}.  In this region the median temperature
detected for the WHIM is about T$\sim$0.4~keV (extending up to 0.5 keV
and also to $<$0.3 keV). The minimum average WHIM emission  (i.e. assuming
the maximum possible subtraction of the LHB contribution) is 
$\sim 240 \times 10^{-6} \,\rm{cts~s^{-1}arcmin^{-2}}$ in the
0.14--0.28~keV ROSAT band, corresponding to a flux 
$\rm F_{0.2-0.4keV}=9\times10^{-16} \,\rm{erg~s^{-1}cm^{-2}arcmin^{-2}}$. 
   \begin{figure*}
   \centering
   \includegraphics[width=0.79\textwidth]{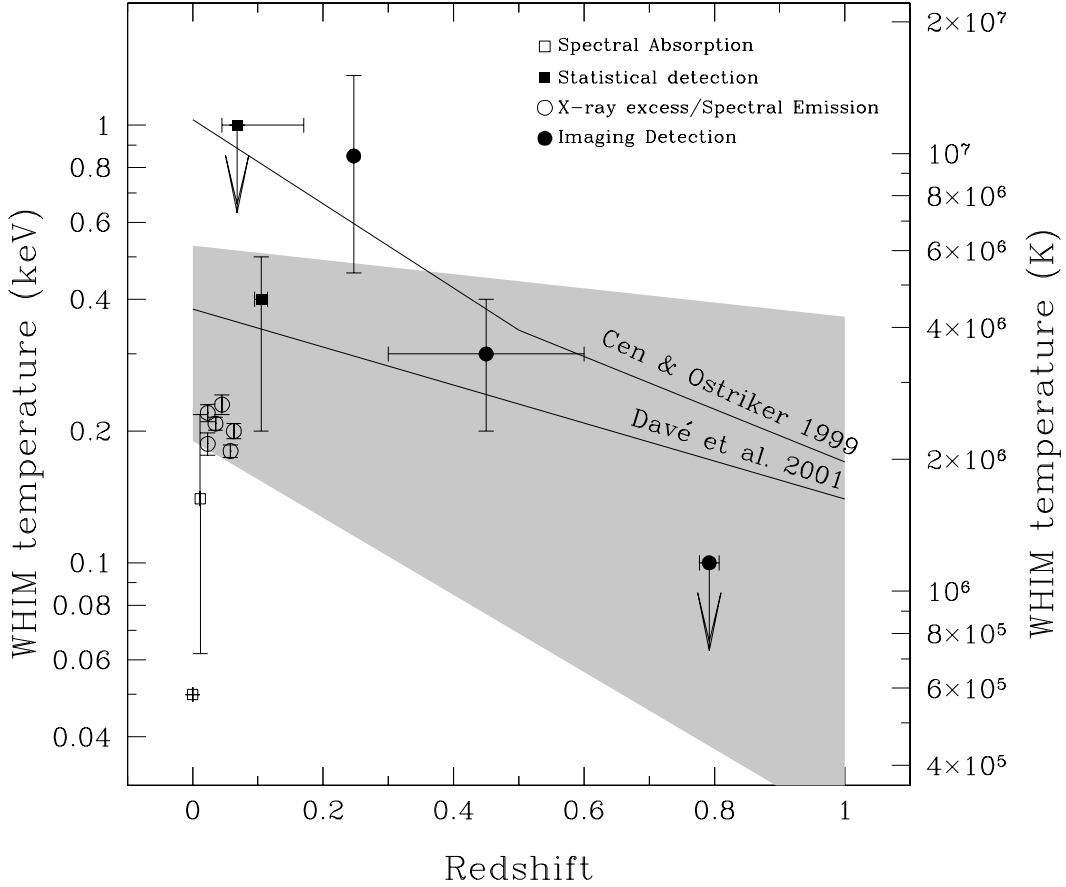}
      \caption{
        Observed WHIM temperatures as a function of the redshift taken from the
        reference reported in Table \ref{tab_temp}. The solid lines 
	are the predictions of equation 4 in \citet{cen} and
        \citet{dave}. The shaded region shows the interval of
	temperatures which includes 
        50\% of the WHIM as 
        reported in Fig.~5 of \citet{dave}.
              }
      \label{temp_hist}
   \end{figure*}
We assume that these values are roughly representative of the emission
by WHIM in the superstructures of the local universe, however we will 
show that our conclusions are not critically dependent on such
assumptions. If we move such
a medium to a redshift of 0.8, in absence of evolution its density
will increase proportionally to (1+z)$^3$ due to the Hubble flow,
and more precisely by a factor [(1+0.8)/(1+0.1)]$^3=$4.4.
However, we also have to include the intrinsic
density evolution expected for the WHIM. From \citet[][
therein Fig.~4]{dave} we can estimate that the WHIM at high redshift (z$\sim$1)
should be denser by a factor of $\sim$1.5 than locally\footnote{These predictions refer to the bulk of the 
WHIM that virtually, having on average
lower densities than the WHIM in supercluster environment, 
should not be detectable with the present day instruments. It is possible that the denser WHIM phase in the 
supercluster environment evolve in a different way. We are not aware of
theoretical works exploring 
the evolution of this dense WHIM phase. }.
When combined with the effect of the Hubble flow, the density increases by a factor
$\rm n_{0.8}/n_{0.1}=6.6$.
The thermal emission increases proportionally to $n^2$, therefore
from z$=$0.1 to z$=$0.8 the WHIM emissivity is expected to
increase by a factor $\rm \epsilon _{0.8}/\epsilon _{0.1} =
(n_{0.8}/n_{0.1})^2=43.6$. Since we are observing (or constraining)
the surface brightness, we have also to account for a cosmological dimming
proportional to (1+z)$^4$, i.e. [(1+0.8)/(1+0.1)]$^4=$7.2. 
Assuming that the temperature does not change (our working 
hypothesis) the spectral shape remains unchanged, but the redshift 
moves the thermal spectrum to lower energies, certainly not enough to 
push the thermal cutoff in our 0.2--0.4~keV band, but the consequent 
\setlength{\extrarowheight}{1pt}
\begin{table*}[t]
  \begin{center}
 \begin{threeparttable}[b]
\caption{WHIM temperatures (sorted by redsfhift) measured so far with
  different methods by various authors.}
    \begin{tabular}{lrrrrr}
\hline
\hline
{\bf Object}		& {\bf z~~}	 & {\bf Object Type}	         & {\bf Method}	& {\bf kT (keV)} & {\bf Reference}\\
\hline
PKS 2155-304    & $0\tnote{a}$ & QSO           	     & Spectral Absorption& $\sim 0.043$           & \citet{nicastro}\\
Mkn 421         & $0.0116\tnote{a}$ & QSO           	     & Spectral Absorption& $0.06\div0.22$    & \citet{nicastro2}\\
Coma            & $0.0231$        & Cluster of Galaxies        & Soft excess   & $0.22\pm0.01\tnote{b}$    & \citet{finoguenov}\\
Coma            & $0.0231$ 	  & Cluster of Galaxies          & Soft excess   & $0.187\pm0.011$  & \citet{kaastra}\\
Abell 2052      & $0.035$ 	  & Cluster of Galaxies          & Soft excess   & $0.208\pm0.007$  & \citet{kaastra}\\
MKW 3s          & $0.045$ 	  & Cluster of Galaxies          & Soft excess   & $0.23\pm0.012$  & \citet{kaastra}\\
%PKS 2155-304    & $0.055\tnote{a}$  & QSO  	     & Spectral Absorption& $0.34\div0.43$      & \citet{fang}\\
Sersic 159 03   & $0.058$ 	  & Cluster of Galaxies          & Soft excess   & $0.18\pm0.006$  & \citet{kaastra}\\
Abell 1795      & $0.0631$ 	  & Cluster of Galaxies          & Soft excess   & $0.2\pm0.008$  & \citet{kaastra}\\
Abell Clusters  & $0.068^{+0.102}_{-0.023}\tnote{c}$  & Clusters of Galaxies      & Statistical  & $< 1$            & \citet{soltan}\\
Sculptor Scl        & $0.105\pm0.01$  & Supercluster     & Statistical  & $0.4^{+0.1}_{-0.2}\tnote{c}$      & \citet{zappacosta2}\\
Abell 2125      & 0.247           & Cluster of Galaxies       & Imaging      & $0.85^{+0.45}_{-0.39}\tnote{d}$   & \citet{wang}\\
Warwick field      & $0.45\pm0.15$            & Field         & Imaging      & $0.3\pm0.15\tnote{c}$     & \citet{zappacosta}\\
Lockman Hole    & $0.791\pm0.016$& Superstructure& Imaging      & $<0.1$           & this paper\\
\hline
    \end{tabular}
  \label{tab_temp}
    \begin{tablenotes}
    \item [a] redshift of the absorption system
    \item [b] $1\sigma$ error-bars
    \item [c] median value
    \item [d] $90$\% level of confidence
%    \item [c] WHIM with lower temperatures may exists
    \end{tablenotes}
 \end{threeparttable}
  \end{center}
\end{table*}
K--correction still changes the observed flux significantly, and more 
specifically by a factor of 1.8. Summarizing all of these effects,
{\it in absence of temperature evolution} the surface
brightness expected for WHIM associated with a superstructure at z$=$0.8
relative to a superstructure at z$=$0.1 is
$$
\rm{{S_{0.2-0.4keV}(0.8) \over {S_{0.2-0.4keV}(0.1)}}
= \left(n_{0.8} \over n_{0.1}\right) ^2 ~
  \left( {1+0.8 \over 1+0.1}\right)^{-4} ~ (1.8)^{-1} = 3.4}
$$
combined with the surface brightness of the WHIM observed in the
Sculptor supercluster, we obtain an expected surface brightness
for the WHIM at z$=$0.8 of
$$\rm  S_{0.2-0.4keV}(0.8)
\approx 3.1 \times 10^{-15} ~erg~s^{-1}~cm^{-2}~arcmin^{-2}$$
 which is roughly an order of magnitude more (more than a factor of 8) than the upper
limit obtained by us for the superstructure in the Lockman Hole. \\
 One possibility could be that filaments are more rare at high
redshift, and that none is present in our field.
However, cosmological simulations predict that filamentary
structures should already be formed by z$\sim$1, and that super-structures
(traced by overdensity of galaxies) are the locations where filaments
are most likely to be present. Since we are clearly investigating one of
such super-structures (as demonstrated by the galaxy overdensity),
the filaments expected by cosmological models at such redshifts are much
more likely to be present in our region than in the field.
Another, more likely possibility
is that baryonic filaments are indeed present in this large
scale structure, but that their temperature at z$=$0.8
is significantly lower than in the local Universe (z$\le$0.1),
making them undetectable.
Indeed a lower temperature moves the thermal cutoff of the spectrum to lower
energies and specifically below the band 0.2-0.4 keV observed by us.  
We have calculated that the maximum allowed temperature at
z=0.8, which
would make the observed flux consistent with our upper limit in the
0.2--0.4~keV band, is about 0.07--0.1~keV. It is important to note that this
temperature limit is weakly sensitive on the density and emissivity
assumptions discussed above. 
Indeed, the important result is that
the expected flux is so much higher than
our upper limit that the only way out
is to require the temperature
to be low enough to move the thermal cutoff below the
0.2--0.4~keV band.

\section{The thermal history of the WHIM} 

A summary of the constraints on the observed WHIM temperatures
as a function of redshift is given in Fig.~\ref{temp_hist}. We include
results by other works listed in Table \ref{tab_temp}
(both emission and absorption detections). 
The points give either the median value or the best fitting value. 
As expected, imaging and statistical measurements, which tend to detect
the brightest WHIM emissions, obtain higher temperature values.
This is a consequence of the WHIM emissivity per
unit energy being proportional to the temperature\footnote{
The thermal emissivity per unit energy is 
related to the 
temperature $T$ and the density $n$ as $T^{-0.5}n^{2}$ (at energies
below the thermal cutoff). 
On the other
hand \citet{dave} showed that the
WHIM temperature and density are directly
proportional, and therefore
$\epsilon \propto T^{1.5}$.}.
On the contrary, absorption systems sample
random regions of the WHIM, including the very low temperature regions.
Soft excess around clusters also tend to be biased against high
temperatures, because the latter would be confused with the
cluster emission itself. Moreover the WHIM origin of 
cluster soft excesses have recently
been partly brought into question by cosmological 
simulations \citep{mittaz2,cheng}.

A homogeneous comparison should be limited to detections
(and upper limits) made with the same technique. In this paper
we have searched for WHIM emission, and therefore our upper limit
must be compared with the other detections and upper limits
obtained with imaging or statistical detections of WHIM emission
(full symbols in Fig.~\ref{temp_hist}).
Such
measurements indicate that the observed WHIM temperature decreases with
redshift (Fig.~~\ref{temp_hist}), just as predicted by cosmological models.
The number of observational data is small and with large errorbars,
however the plot in Fig.~\ref{temp_hist} is the first attempt
of constraining the evolution of the WHIM temperature with redshift,
with the currently available data.

It is possible to go beyond this qualitative statement and investigate the
quantitative agreement with the model predictions. 
\citet{cen} presented in their paper a simple argument that can reproduce
representative values for the temperature of the WHIM 
\citep[see also Fig. 5 in][]{dave}. 
They infer the postshock temperature of a cosmic gas collapsing inside
a slightly nonlinear large-scale structure of size L as $T \propto
c_s^{2} \approx K(Lt_H^{-1}(z))^{2}$, where $c_s$ is the gas sound speed and
$t_H(z)$ the Hubble time at the redshift of interest and K is constant  
\citep[see equation 4 in][]{cen}.
%They infer the postshock temperature
%of an infalling gas estimating the speed of sound 
%\citep[equation 4 in][]{cen} behind the shock as 
%the time occurred for a perturbation to collapse inside a given
%slightly non linear large-scale structure.
Such simple estimate is shown to correctly reproduce the
results from numerical simulations \citep{cen,dave}.
The same approach was used by \citet{dave} who obtain similar
results.
In Fig.~\ref{temp_hist} we show the trends of the WHIM temperature
as a function of redshift obtained by \citet{cen} and \citet{dave} by using the
argument discussed above (the shaded area indicates the interval
of temperatures which includes $\sim$50\% of the WHIM in the
\citet{dave} distributions).
%of the peak of the distributions of WHIM at various redshifts.
%we have reported the prediction of
%this argument that in \citet{cen} (long-dashed line in their Fig. 1) 
% reproduce the density--weighted temperature of their simulation 
%and in \citet{dave} fairly fit the peak temperatures of their Fig. 5.
Although the uncertainties of the theoretical models are large,
both predictions fit reasonably well the data points, and
in particular the temperatures inferred from the imaging and statistical
methods.

\vskip1cm

\section{Conclusions}
We have identified a large-scale structure of galaxies in the Lockman
Hole at redshift z~$\sim0.79\pm0.015$ by means of an optical 
spectroscopic survey. 
The superstructure extends over a
region of more than 7.5~Mpc (in projection) and is structured 
in three sub-concentrations at median redshifts of 0.776, 0.784 and
0.806. In this superstructure the WHIM
predicted by cosmological models should have already formed. 
By analysing ROSAT and XMM pointings we could set a tight upper
limit on the WHIM emission associated with the superstructure. 
From this flux limit we could estimate 
an upper limit of $\sim$0.1~keV on the WHIM
temperature at z$\sim$0.8.
The combination of this tight upper limit with other previous WHIM temperature
measurements (at lower redshifts) strongly suggests that the
WHIM temperature must be rapidly decreasing with redshift,
as expected by the cosmological models. 
The agreement of the redshift distribution of the
observed WHIM temperatures with the cosmological
predictions \citep{cen,dave} is
reasonably good even from a quantitative point of view.
However further work is required to improve the statistics
on the WHIM temperature measurements (or constraints)
at high redshift.

\begin{acknowledgements}

The paper is based on observations obtained with XMM-Newton, an ESA
science mission with instruments and contributions directly funded by
ESA Member States and the USA (NASA). The XMM-Newton project is
supported by the Bundesministerium f\"{u}r Bildung und
Forschung/Deutsches Zentrum f\"{u}r Luft- und Raumfahrt (BMFT/DLR),
the Max-Planck Society and the Heidenhain-Stiftung, and also by PPARC,
CEA, CNES, and ASI. This work was partially supported by the Italian
Ministry of Research (MIUR) and by the Italian Institute of
Astrophysics (INAF). We thank G. Hasinger, H. Boehringer, X. Barcons
and A. Fabian for letting us to use the Lockman Hole image to remove
point source contamination and Konrad Dennerl for his assistance with
XMM data analysis.  AF acknowledges support from BMBF/DLR under grant
50 OR 0207 and MPG.

      \end{acknowledgements}

\bibliography{zappacosta}

\pagebreak

\appendix

\section{Catalog of Objects}\label{catalog}

\setlength{\extrarowheight}{3pt}
  \begin{center}
\tablehead{%
\hline
\hline
 {\bf Id} & {\bf RA (J2000)} & {\bf DEC (J2000)} & {\bf z} & {\bf Quality} \\
\hline
}
\tabletail{%
\hline
}
 \begin{mpxtabular}{rcccc}
     1	  & 10$^h$53$^m$41.77$^s$ & +57$^{\circ}$21$^{\prime}$29.56$^{\prime\prime}$ & 0.281 & low \\
     2	  & 10$^h$53$^m$45.41$^s$ & +57$^{\circ}$21$^{\prime}$40.96$^{\prime\prime}$ & 0.085 & high \\
     3	  & 10$^h$53$^m$43.71$^s$ & +57$^{\circ}$22$^{\prime}$01.01$^{\prime\prime}$ & 0.608 & low \\
     4	  & 10$^h$53$^m$28.84$^s$ & +57$^{\circ}$22$^{\prime}$05.78$^{\prime\prime}$ & 0.776 & low \\
     5	  & 10$^h$53$^m$36.44$^s$ & +57$^{\circ}$22$^{\prime}$17.45$^{\prime\prime}$ & 0.482 & low \\
     6	  & 10$^h$53$^m$33.12$^s$ & +57$^{\circ}$22$^{\prime}$34.99$^{\prime\prime}$ & 0.735 & high \\
     7	  & 10$^h$53$^m$30.17$^s$ & +57$^{\circ}$22$^{\prime}$41.48$^{\prime\prime}$ & 0.863 & low \\
     8	  & 10$^h$53$^m$30.86$^s$ & +57$^{\circ}$22$^{\prime}$48.10$^{\prime\prime}$ & 0.634 & low \\
     9	  & 10$^h$53$^m$19.95$^s$ & +57$^{\circ}$22$^{\prime}$51.48$^{\prime\prime}$ & 0.528 & low \\
    10	  & 10$^h$53$^m$34.39$^s$ & +57$^{\circ}$23$^{\prime}$06.16$^{\prime\prime}$ & 0.405 & low \\
    11	  & 10$^h$53$^m$38.38$^s$ & +57$^{\circ}$23$^{\prime}$06.04$^{\prime\prime}$ & 0.419 & high \\
    12	  & 10$^h$53$^m$44.53$^s$ & +57$^{\circ}$23$^{\prime}$26.43$^{\prime\prime}$ & 1.017 & high \\
    13	  & 10$^h$53$^m$37.79$^s$ & +57$^{\circ}$23$^{\prime}$28.09$^{\prime\prime}$ & 0.436 & low \\
    14	  & 10$^h$53$^m$41.86$^s$ & +57$^{\circ}$23$^{\prime}$53.08$^{\prime\prime}$ & 0.892 & low \\
    15	  & 10$^h$53$^m$45.87$^s$ & +57$^{\circ}$24$^{\prime}$10.48$^{\prime\prime}$ & 0.777 & high \\
    16	  & 10$^h$53$^m$38.60$^s$ & +57$^{\circ}$24$^{\prime}$08.85$^{\prime\prime}$ & 0.484 & high \\
    17	  & 10$^h$53$^m$47.71$^s$ & +57$^{\circ}$27$^{\prime}$15.48$^{\prime\prime}$ & 0.785 & high \\
    18	  & 10$^h$53$^m$43.83$^s$ & +57$^{\circ}$26$^{\prime}$59.82$^{\prime\prime}$ & 0.729 & low \\
    19	  & 10$^h$53$^m$54.44$^s$ & +57$^{\circ}$26$^{\prime}$11.21$^{\prime\prime}$ & 0.644 & low \\
    20	  & 10$^h$53$^m$40.40$^s$ & +57$^{\circ}$26$^{\prime}$36.23$^{\prime\prime}$ & 0.821 & low \\
    21	  & 10$^h$53$^m$45.22$^s$ & +57$^{\circ}$26$^{\prime}$25.45$^{\prime\prime}$ & 0.808 & high \\
    22	  & 10$^h$53$^m$26.84$^s$ & +57$^{\circ}$26$^{\prime}$30.70$^{\prime\prime}$ & 0.354 & high \\
    23	  & 10$^h$53$^m$45.09$^s$ & +57$^{\circ}$25$^{\prime}$56.43$^{\prime\prime}$ & 0.783 & high \\
    24	  & 10$^h$53$^m$23.87$^s$ & +57$^{\circ}$27$^{\prime}$29.10$^{\prime\prime}$ & 0.807 & low \\
    25	  & 10$^h$53$^m$44.91$^s$ & +57$^{\circ}$26$^{\prime}$03.48$^{\prime\prime}$ & 0.619 & high \\
    26	  & 10$^h$53$^m$39.80$^s$ & +57$^{\circ}$25$^{\prime}$05.50$^{\prime\prime}$ & 0.807 & high \\
    27	  & 10$^h$53$^m$39.41$^s$ & +57$^{\circ}$25$^{\prime}$30.62$^{\prime\prime}$ & 0.335 & high \\
    28	  & 10$^h$53$^m$44.89$^s$ & +57$^{\circ}$27$^{\prime}$27.74$^{\prime\prime}$ & 0.594 & high \\
    29	  & 10$^h$53$^m$32.13$^s$ & +57$^{\circ}$25$^{\prime}$07.74$^{\prime\prime}$ & 0.479 & high \\
    30	  & 10$^h$53$^m$44.77$^s$ & +57$^{\circ}$24$^{\prime}$45.67$^{\prime\prime}$ & 0.917 & low \\
    31	  & 10$^h$53$^m$27.98$^s$ & +57$^{\circ}$27$^{\prime}$53.20$^{\prime\prime}$ & 0.146 & low \\
    32	  & 10$^h$53$^m$42.73$^s$ & +57$^{\circ}$27$^{\prime}$51.53$^{\prime\prime}$ & 0.684 & high \\
    33	  & 10$^h$53$^m$48.68$^s$ & +57$^{\circ}$28$^{\prime}$33.26$^{\prime\prime}$ & 0.279 & low \\
    34	  & 10$^h$53$^m$58.52$^s$ & +57$^{\circ}$28$^{\prime}$37.64$^{\prime\prime}$ & 0.751 & low \\
    35	  & 10$^h$53$^m$30.46$^s$ & +57$^{\circ}$28$^{\prime}$35.05$^{\prime\prime}$ & 0.531 & low \\
    36	  & 10$^h$53$^m$30.22$^s$ & +57$^{\circ}$28$^{\prime}$55.08$^{\prime\prime}$ & 1.011 & low \\
    37	  & 10$^h$53$^m$48.84$^s$ & +57$^{\circ}$29$^{\prime}$17.08$^{\prime\prime}$ & 0.889 & high \\
    38	  & 10$^h$53$^m$56.29$^s$ & +57$^{\circ}$29$^{\prime}$25.65$^{\prime\prime}$ & 0.580 & high \\
    39	  & 10$^h$53$^m$50.98$^s$ & +57$^{\circ}$29$^{\prime}$33.65$^{\prime\prime}$ & 0.776 & low \\
    40	  & 10$^h$53$^m$40.91$^s$ & +57$^{\circ}$29$^{\prime}$42.15$^{\prime\prime}$ & 0.919 & low \\
    41	  & 10$^h$53$^m$44.08$^s$ & +57$^{\circ}$30$^{\prime}$17.99$^{\prime\prime}$ & 0.490 & low \\
    42	  & 10$^h$53$^m$26.96$^s$ & +57$^{\circ}$31$^{\prime}$10.83$^{\prime\prime}$ & 0.377 & low \\
    43	  & 10$^h$53$^m$56.19$^s$ & +57$^{\circ}$31$^{\prime}$14.36$^{\prime\prime}$ & 0.701 & low \\
    44	  & 10$^h$53$^m$32.85$^s$ & +57$^{\circ}$31$^{\prime}$52.09$^{\prime\prime}$ & 0.422 & high \\
    45	  & 10$^h$53$^m$48.02$^s$ & +57$^{\circ}$32$^{\prime}$05.64$^{\prime\prime}$ & 0.783 & low \\
    46	  & 10$^h$53$^m$40.31$^s$ & +57$^{\circ}$32$^{\prime}$09.61$^{\prime\prime}$ & 0.692 & low \\
    47	  & 10$^h$53$^m$42.17$^s$ & +57$^{\circ}$32$^{\prime}$25.60$^{\prime\prime}$ & 0.321 & high \\
    48	  & 10$^h$53$^m$49.23$^s$ & +57$^{\circ}$35$^{\prime}$39.75$^{\prime\prime}$ & 0.231 & low \\
    49	  & 10$^h$53$^m$49.31$^s$ & +57$^{\circ}$34$^{\prime}$28.74$^{\prime\prime}$ & 0.391 & low \\
    50	  & 10$^h$53$^m$40.12$^s$ & +57$^{\circ}$36$^{\prime}$19.28$^{\prime\prime}$ & 0.784 & high \\
    51	  & 10$^h$53$^m$33.68$^s$ & +57$^{\circ}$36$^{\prime}$10.40$^{\prime\prime}$ & 0.500 & high \\
    52	  & 10$^h$53$^m$37.67$^s$ & +57$^{\circ}$36$^{\prime}$04.33$^{\prime\prime}$ & 0.405 & low \\
    53	  & 10$^h$53$^m$40.23$^s$ & +57$^{\circ}$35$^{\prime}$50.81$^{\prime\prime}$ & 0.933 & low \\
    54	  & 10$^h$53$^m$44.73$^s$ & +57$^{\circ}$36$^{\prime}$08.73$^{\prime\prime}$ & 0.591 & low \\
    55	  & 10$^h$54$^m$01.68$^s$ & +57$^{\circ}$35$^{\prime}$34.23$^{\prime\prime}$ & 0.256 & low \\
    56	  & 10$^h$53$^m$41.03$^s$ & +57$^{\circ}$35$^{\prime}$12.77$^{\prime\prime}$ & 0.493 & low \\
    57	  & 10$^h$53$^m$44.76$^s$ & +57$^{\circ}$35$^{\prime}$15.24$^{\prime\prime}$ & 0.093 & low \\
    58	  & 10$^h$53$^m$42.62$^s$ & +57$^{\circ}$34$^{\prime}$03.58$^{\prime\prime}$ & 0.784 & high \\
    59	  & 10$^h$53$^m$41.48$^s$ & +57$^{\circ}$34$^{\prime}$28.55$^{\prime\prime}$ & 0.256 & low \\
    60	  & 10$^h$53$^m$46.09$^s$ & +57$^{\circ}$34$^{\prime}$11.09$^{\prime\prime}$ & 1.084 & low \\
    61	  & 10$^h$53$^m$42.87$^s$ & +57$^{\circ}$34$^{\prime}$03.80$^{\prime\prime}$ & 0.784 & high \\
    62	  & 10$^h$53$^m$36.94$^s$ & +57$^{\circ}$33$^{\prime}$57.55$^{\prime\prime}$ & 0.420 & high \\
    63	  & 10$^h$53$^m$36.32$^s$ & +57$^{\circ}$33$^{\prime}$13.74$^{\prime\prime}$ & 0.706 & high \\
    64	  & 10$^h$53$^m$36.32$^s$ & +57$^{\circ}$33$^{\prime}$12.91$^{\prime\prime}$ & 0.784 & high \\
    65	  & 10$^h$53$^m$45.19$^s$ & +57$^{\circ}$33$^{\prime}$42.05$^{\prime\prime}$ & 0.701 & high \\
    66	  & 10$^h$54$^m$03.72$^s$ & +57$^{\circ}$33$^{\prime}$29.03$^{\prime\prime}$ & 0.662 & low \\
    67	  & 10$^h$53$^m$37.33$^s$ & +57$^{\circ}$32$^{\prime}$52.90$^{\prime\prime}$ & 0.562 & low \\
    68	  & 10$^h$53$^m$38.25$^s$ & +57$^{\circ}$32$^{\prime}$49.74$^{\prime\prime}$ & 0.785 & high \\
    69	  & 10$^h$53$^m$29.13$^s$ & +57$^{\circ}$36$^{\prime}$41.92$^{\prime\prime}$ & 0.427 & high \\
    70	  & 10$^h$52$^m$50.23$^s$ & +57$^{\circ}$28$^{\prime}$25.09$^{\prime\prime}$ & 0.536 & low \\
    71	  & 10$^h$52$^m$49.71$^s$ & +57$^{\circ}$28$^{\prime}$41.56$^{\prime\prime}$ & 0.342 & high \\
    72	  & 10$^h$52$^m$39.67$^s$ & +57$^{\circ}$28$^{\prime}$57.53$^{\prime\prime}$ & 0.669 & low \\
    73	  & 10$^h$52$^m$48.20$^s$ & +57$^{\circ}$29$^{\prime}$54.99$^{\prime\prime}$ & 0.291 & low \\
    74	  & 10$^h$52$^m$41.79$^s$ & +57$^{\circ}$30$^{\prime}$53.26$^{\prime\prime}$ & 0.206 & high \\
    75	  & 10$^h$52$^m$42.14$^s$ & +57$^{\circ}$31$^{\prime}$23.02$^{\prime\prime}$ & 0.121 & high \\
    76	  & 10$^h$52$^m$55.15$^s$ & +57$^{\circ}$31$^{\prime}$47.06$^{\prime\prime}$ & 0.343 & low \\
    77	  & 10$^h$52$^m$53.80$^s$ & +57$^{\circ}$33$^{\prime}$11.11$^{\prime\prime}$ & 0.669 & high \\
    78	  & 10$^h$52$^m$45.50$^s$ & +57$^{\circ}$28$^{\prime}$05.49$^{\prime\prime}$ & 0.606 & high \\
    79	  & 10$^h$52$^m$41.35$^s$ & +57$^{\circ}$27$^{\prime}$21.95$^{\prime\prime}$ & 0.552 & high \\
    80	  & 10$^h$52$^m$49.02$^s$ & +57$^{\circ}$26$^{\prime}$39.76$^{\prime\prime}$ & 0.322 & low \\
    81	  & 10$^h$52$^m$42.18$^s$ & +57$^{\circ}$26$^{\prime}$02.02$^{\prime\prime}$ & 0.343 & high \\
    82	  & 10$^h$54$^m$20.31$^s$ & +57$^{\circ}$24$^{\prime}$02.86$^{\prime\prime}$ & 0.778 & low \\
    83	  & 10$^h$54$^m$29.59$^s$ & +57$^{\circ}$24$^{\prime}$11.82$^{\prime\prime}$ & 0.776 & high \\
    84	  & 10$^h$54$^m$23.22$^s$ & +57$^{\circ}$24$^{\prime}$46.59$^{\prime\prime}$ & 0.809 & low \\
    85	  & 10$^h$54$^m$17.46$^s$ & +57$^{\circ}$25$^{\prime}$02.44$^{\prime\prime}$ & 0.138 & high \\
    86	  & 10$^h$54$^m$27.92$^s$ & +57$^{\circ}$26$^{\prime}$14.55$^{\prime\prime}$ & 0.578 & high \\
    87	  & 10$^h$54$^m$14.04$^s$ & +57$^{\circ}$27$^{\prime}$24.94$^{\prime\prime}$ & 0.913 & low \\
    88	  & 10$^h$54$^m$29.28$^s$ & +57$^{\circ}$27$^{\prime}$44.63$^{\prime\prime}$ & 0.306 & low \\
    89	  & 10$^h$54$^m$11.02$^s$ & +57$^{\circ}$31$^{\prime}$42.79$^{\prime\prime}$ & 0.396 & high \\
    90	  & 10$^h$54$^m$22.84$^s$ & +57$^{\circ}$31$^{\prime}$12.17$^{\prime\prime}$ & 0.343 & high \\
    91	  & 10$^h$54$^m$05.16$^s$ & +57$^{\circ}$31$^{\prime}$20.91$^{\prime\prime}$ & 0.977 & low \\
    92	  & 10$^h$54$^m$05.16$^s$ & +57$^{\circ}$31$^{\prime}$22.56$^{\prime\prime}$ & 1.062 & low \\
    93	  & 10$^h$54$^m$10.99$^s$ & +57$^{\circ}$30$^{\prime}$56.14$^{\prime\prime}$ & 0.226 & low \\
    94	  & 10$^h$54$^m$30.51$^s$ & +57$^{\circ}$29$^{\prime}$57.57$^{\prime\prime}$ & 0.619 & high \\
    95	  & 10$^h$54$^m$01.59$^s$ & +57$^{\circ}$28$^{\prime}$13.58$^{\prime\prime}$ & 0.890 & low \\
    96	  & 10$^h$54$^m$03.72$^s$ & +57$^{\circ}$29$^{\prime}$15.01$^{\prime\prime}$ & 0.806 & high \\
    97	  & 10$^h$54$^m$23.93$^s$ & +57$^{\circ}$29$^{\prime}$10.58$^{\prime\prime}$ & 0.410 & low \\
    98	  & 10$^h$54$^m$16.76$^s$ & +57$^{\circ}$29$^{\prime}$03.93$^{\prime\prime}$ & 0.589 & low \\
    99	  & 10$^h$54$^m$03.14$^s$ & +57$^{\circ}$28$^{\prime}$57.45$^{\prime\prime}$ & 0.793 & low \\
   100	  & 10$^h$54$^m$22.93$^s$ & +57$^{\circ}$28$^{\prime}$42.62$^{\prime\prime}$ & 0.787 & high \\
   101	  & 10$^h$54$^m$29.85$^s$ & +57$^{\circ}$28$^{\prime}$17.79$^{\prime\prime}$ & 0.711 & high \\
   102	  & 10$^h$54$^m$21.11$^s$ & +57$^{\circ}$30$^{\prime}$14.17$^{\prime\prime}$ & 0.965 & low \\
   103	  & 10$^h$54$^m$33.97$^s$ & +57$^{\circ}$29$^{\prime}$52.14$^{\prime\prime}$ & 0.276 & high \\
\hline
     \end{mpxtabular}
\end{center}

\end{document}